\documentclass{workshop}
\begin{document}

\title{
Glowbug, a Low-Cost, High-Sensitivity Gamma-Ray Burst Telescope\\ 
}

\author{
J.E. Grove,$^1$ C.C. Cheung,$^1$ M. Kerr,$^1$ L.J. Mitchell,$^1$ B.F. Phlips,$^1$ R.S. Woolf,$^1$ E.A. Wulf,$^1$ \\
M.S. Briggs,$^2$ C.A. Wilson-Hodge,$^3$ D. Kocevski,$^3$ and J. Perkins$^4$
\\[12pt]  
%
$^1$  U.S. Naval Research Laboratory, Washington, DC \\
$^2$  University of Alabama in Huntsville, Huntsville, AL \\
$^3$  NASA Marshall Space Flight Center, Huntsville, AL \\
$^4$  NASA Goddard Space Flight Center, Greenbelt, MD \\
%
\textit{E-mail: eric.grove@nrl.navy.mil} 
}

\abst{
We describe Glowbug, a gamma-ray telescope for bursts and other transients in the 30 keV to 2 MeV band. It was recently selected for funding by the NASA Astrophysics Research and Analysis program, with an expected launch in the early 2020s. Similar in concept to the Fermi Gamma Burst Monitor (GBM) and with similar sensitivity, Glowbug will join and enhance future networks of burst telescopes to increase sky coverage to short Gamma-Ray Bursts (SGRBs) from binary neutron star (BNS) mergers, including possible SGRBs from NS-black hole mergers. With the recent discovery of the SGRB coincident with the gravitational wave transient GW170817, we know such events occur with reasonable frequency. Expanded sky coverage in gamma rays is essential, as more detections of gravitational waves are expected with the improved sensitivity of the upgraded ground-based interferometers in the coming years.
}

\kword{workshop: proceedings --- gamma-ray bursts --- instrumentation}

\maketitle
\thispagestyle{empty}

\section{Introduction}
Gamma-Ray Bursts (GRBs) are the most luminous transient events in the Universe and were discovered during the Vela satellite programs in the 1960s. They were determined to be of cosmic origin based on their individual directions \citep{Kle73,Cli73} and extragalactic in nature from the discovery of host galaxies enabled by precise localization of X-ray afterglow \citep{Pir99}. GRBs can be separated observationally into two classes \citep{Kou93} based on durations of less than 2s (short GRBs) and $>$2s (long GRBs). After decades of study, the emergent picture is that these sub-classes originate from compact binary mergers and the collapse of massive stars, respectively.

Glowbug is a  gamma-ray telescope for GRB detection similar in concept to the Fermi GBM \citep{Meegan2009}.  The design is simple, with an assembly of relatively large scintillator panels arrayed on a half cube.  This array provides full-sky coverage with good sensitivity, as well as modest localization ability.  Scintillator thicknesses are chosen to cover the hard X-ray to soft gamma-ray band where the peak of the GRB emission spectrum lies.  Instrument development is funded by NASA, while the DoD Space Test Program (STP) will provide launch and one-year of on-orbit mission operations.  We are currently targeting an STP launch opportunity to the International Space Station in 2023.

Glowbug will perform forefront GRB research, detecting an estimated hundreds of long GRBs and dozens of short GRBs per year, and providing burst spectra, lightcurves, and positions.  Glowbug’s primary science objective is the detection and localization of short GRBs, which are the result of mergers of compact binaries \citep{Abbott2017a,Abbott2017b} involving a neutron star with either another neutron star (NS-NS) or a black hole (NS-BH). The instrument is designed to complement existing (but mature) GRB detection systems (Fermi GBM and Swift BAT), and it will join future networks of small GRB instruments, to provide all-sky coverage and improved localization of such events. Of immediate importance are the BNS systems within the gravitational-wave detection horizon of ~200 Mpc expected from the planned upgrades to the Advanced LIGO-Virgo detectors in the early 2020s \citep{Abbott2016}. 

\section{Instrument Design}
A low-cost GRB detector and localizer can be designed and built using mature technologies.  Glowbug is designed to cover the energy range from $\sim$30 keV to $\sim$2 MeV with a wide field of view (8 sr) and good flux sensitivity to maximize detection probability for fast transients.  The segmented design of Glowbug allows reconstruction of the burst location with modest accuracy.  We have chosen to emphasize detection sensitivity over GRB localization in the design of the instrument.

\begin{figure}[hbt]
\centering
\includegraphics[width=0.5\textwidth]{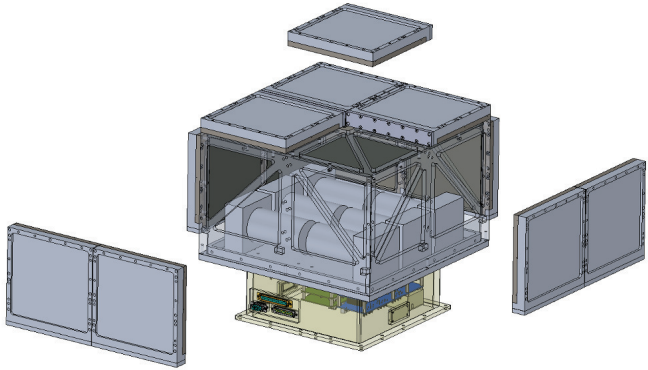}
\caption{Exploded view of Glowbug showing an array of 12 CsI(Tl) detector panels surrounding six CLLB scintillators. The CsI panels provide the primary GRB detection and localization ability, while the CLLB detectors add to the effective area above $\sim$1 MeV.}
\label{fig:explodedView}
\end{figure}

Glowbug relies extensively on experience and heritage from the Strontium Iodide Radiation Instrumentation (SIRI) series of gamma-ray instruments built at NRL to space-qualify a new scintillator material (not used in Glowbug) and silicon photomultiplier (SiPM) photo-sensors for readout.  SIRI-1 \citep{Mit17} successfully completed its one-year low-Earth-orbit mission in December 2019, demonstrating SiPM performance in the LEO radiation environment \citep{Mit19b}.  SIRI-2 \citep{Mit19a} has completed environmental testing, is integrated with its host spacecraft, and is awaiting launch in late 2020.  Importantly, Glowbug will also use the SIRI-2 data acquisition system design, including front-end electronics, detector bias power, and instrument processor.  Glowbug will use the same SensL J-series SiPM as both SIRI instruments, albeit on a different custom carrier board.  We have chosen to use SiPM photosensors because they provide a low size, weight, and power read out for scintillators, and with the success of SIRI-1 the SensL devices are space-qualified. Overall, exploiting heritage from SIRI decreases cost and schedule risk, and increases reliability.

An exploded view of Glowbug is shown in Figure \ref{fig:explodedView}.  The 12 CsI(Tl) panels, each 15$\times$15$\times$1 cm$^3$ and read out by a custom array of SensL J-series 6-mm SiPMs, are mounted on the surface of a half-cube.  Each panel views the sky through a thin aluminum window and is backed by a steel and tantalum graded shield to provide strong attenuation below $\sim$300 keV.  Each panel has approximately cosine angular response in its forward hemisphere, with the forward/backward symmetry broken by the steel and tantalum back shield.  We chose CsI(Tl) for its high density ($\rho = 4.51$ g/cm$^3$) and good gamma-ray stopping power, as well as because it is only very mildly hygroscopic.  This greatly simplifies crystal handling and enclosure design, and it eliminates the need for a hermetic enclosure with optical window, allowing the SiPM array to be closely coupled to the scintillator, improving the scintillation light collection, and lowering the energy threshold.

Within the half-cube array of CsI panels is a set of six Cs$_{2}$LiLaBr$_{6}$(Ce) cylindrical scintillators, each 5 cm diameter $\times$ 10 cm in length, read out at one end by a custom array of SensL J-series SiPMs.  The CLLB detectors provide additional effective area above $\sim$1 MeV for detection and spectral characterization.  CLLB has excellent spectral resolution ($<$4$\%$ FWHM at 662 keV) and is sensitive to thermal neutrons.  Neither of these properties is helpful for GRB science, but a secondary goal of Glowbug is to qualify this new scintillator for use in space and study its activation in LEO.

The CsI and CLLB scintillator arrays are mounted on top of the housing for the instrument central electronics, which acts as a riser to elevate the sensors above surrounding material on the space vehicle (or ISS platform), providing a clear sky view.  Glowbug's central electronics system is derived directly from that of SIRI-2.  It includes power conditioning and conversion, data acquisition and communications via a commercial off-the-shelf (COTS) single-board computer (SBC), and precise timing from GPS.  Each of the 18 detectors has an independent, programmable bias voltage supply, and an independent analog and digital COTS multi-channel analyzer (MCA) to provide time-tagged event-by-event readout of the detector or time-averaged histograms of spectral events.  To ensure precise and accurate event time stamps, each COTS MCA accepts a one pulse-per-second GPS time tone.  Event-list or rate data are aggregated by the SBC and formed into a low-level science data product for on-board analysis and telemetry to the ground.

Glowbug will autonomously detect GRBs in real time following algorithms developed and optimized for the Fermi GBM.  Instrument flight software will generate a Burst Trigger when detector count rates over selected energy ranges exceed a configurable signal-to-noise threshold, which can reconfigure the instrument to downlink event-list data rather than continuous count rate data (which might be required by telemetry downlink budgets).  On-board burst detection will also initiate a Burst Alert message for rapid downlink and distribution into a network such as the Gamma-ray Coordinates Network (GCN) for near real-time sharing of GRB data.

\begin{figure}[hbt]
\centering
\includegraphics[width=0.45\textwidth]{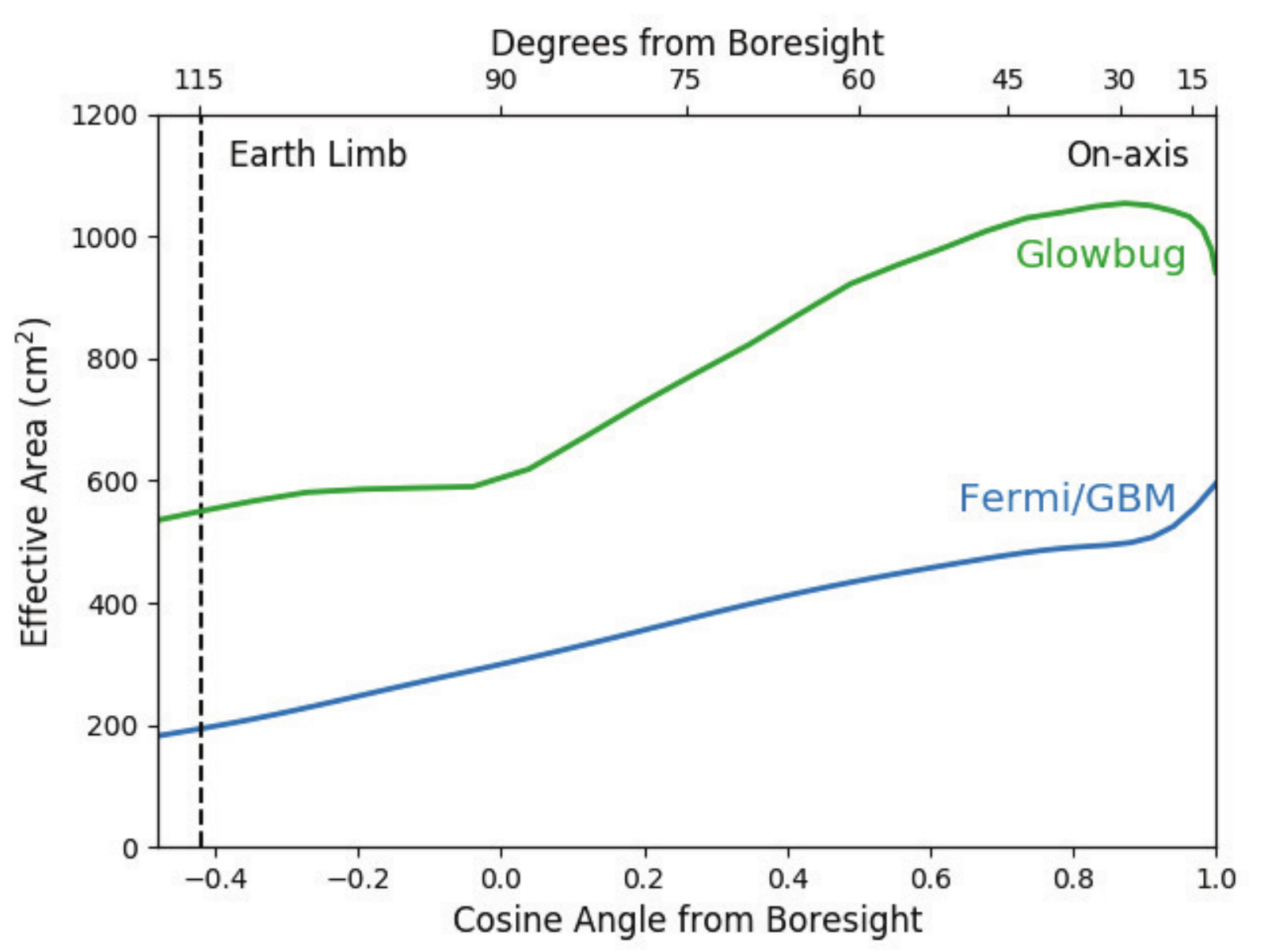}
\caption{Estimated azimuth-averaged effective area for the 12 Glowbug CsI detectors (green) and a toy model of the 12 Fermi GBM detectors (blue) in response to a typical SGRB spectrum.  The effective area of Glowbug is approximately twice that of Fermi GBM.}
\label{fig:EffectiveArea}
\end{figure}

\section{Instrument Performance}
To estimate the performance of the instrument design, we generated a detailed Monte Carlo simulation of the detector modules and the instrument geometry and processed the results into an instrument response matrix.  For the purposes of performance estimates, we ignore counts in the CLLB detectors.  Figure \ref{fig:EffectiveArea} shows the resulting effective area as a function of angle from the zenith.  For comparison, we generated a toy model of Fermi GBM approximating the response and shielding of its 12 detector modules by the Fermi spacecraft and LAT.  Glowbug provides about twice the effective area, in general agreement with the respective surface areas of the detector modules.

We characterized the sensitivity to bursts by developing maximum likelihood algorithms for detection and reconstruction and applying these to simulated data.  Background estimates were derived by scaling Fermi GBM spectra.  We assessed localization performance by applying these algorithms to simulated short GRB matching the properties of those observed by Fermi GBM.  We find typical localization (68\% statistical uncertainty) $<5^{\circ}$, compared to $\sim10^{\circ}$ for GBM --- although the systematic uncertainties depend strongly on the amount of nearby scattering material.

\begin{figure}[hbt]
\centering
\includegraphics[width=0.45\textwidth]{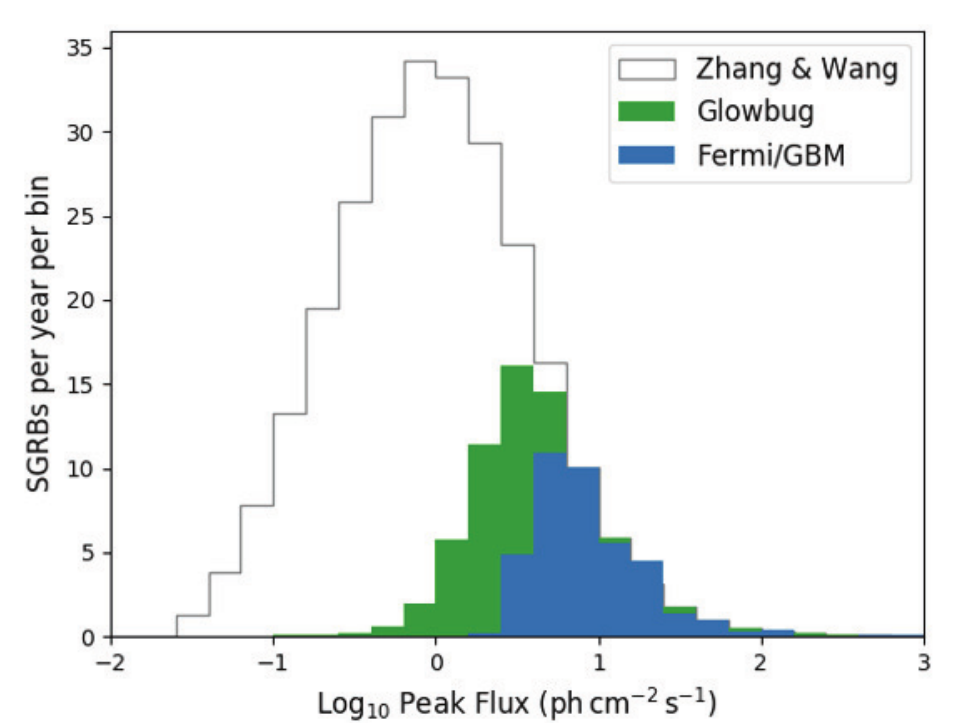}
\caption{Flux distribution for SGRBs from \citet{ZhangWang2018} as a function of peak gamma-ray flux (50-300 keV).  SGRBs detectable by Fermi GBM (blue) and Glowbug (green, including the area underlying the blue histogram) at $>$5-sigma significance are indicated.}
\label{fig:SGRBpeakFluxDistro}
\end{figure}

The increase of about a factor of two in effective area relative to Fermi GBM expands the horizon for faint sources in the local universe by a factor of $\sim$1.4, and thus the volume by a factor of $\sim$3.  Using a limiting fluence from the above performance assessment and applying a luminosity function for SGRBs from the literature \citep{ZhangWang2018}  yields the number detectable at $>$5-sigma significance (Figure \ref{fig:SGRBpeakFluxDistro}).  From this analysis we estimate that Glowbug will detect $\sim$70 SGRBs per year, approximately twice the rate detected by Fermi GBM.  We expect that this factor-of-two increase in SGRB detection rate is fairly robust for a range of reasonable luminosity functions.  

\section{Conclusion}
With a launch in the early 2020s and a one-year nominal mission, Glowbug will be a technology demonstrator for future networks of sensitive, low-cost gamma-ray transient detectors as secondary payloads on larger missions or primary payloads on SmallSat (i.e. larger than CubeSat) missions, providing gamma-ray context for multiwavelength and multimessenger studies of GRBs and other energetic transients.

\section{Acknowledgements}
Glowbug is supported by NASA under the Astrophysics Research and Analysis program.  Initial design work and adaptation of designs from heritage instruments at NRL was supported by the Chief of Naval Research.

\label{last}

\end{document}